\documentclass[aps,prd,showkeys,showpacs,amssymb,
amsfonts,epsf,preprintnumbers,nofootinbib,superscriptaddress]{revtex4}

\usepackage[dvips]{graphicx}
\usepackage{bm,latexsym,amsmath,amssymb,amsfonts,color} 

\newcommand{\nn}{\nonumber}

\begin{document}

\title{Physics at the surface of a star in Eddington-inspired Born-Infeld gravity}

\author{Hyeong-Chan Kim}
\email{hckim@ut.ac.kr}
\affiliation{School of Liberal Arts and Sciences, Korea National University of Transportation, Chungju 380-702, Korea}

\begin{abstract}
We study phenomena happening at the surface of a star in Eddington-inspired Born-Infeld (EiBI) gravity.
The star is made of particles, which are effectively described by a polytropic fluid.
The EiBI theory was known to have a pathology that singularities happen at a star surface.
We suggest that the gravitational back-reaction on the particles cures the problem.  Strong tidal forces near the (surface) singularity modify the effective equation of state of the particles or make the surface be unstable depending on its matter contents.
The geodesic deviation equations take after the Hooke's law, where its frequency-squared is proportional to the scalar curvature at the surface.
For a positive curvature, a particle collides with a probing wall more often and increases the pressure.
With the increased pressure, the surface is no longer singular.
For a negative curvature, the matters around the surface experience repulsions with infinite accelerations.
Therefore, the EiBI gravity is saved from the pathology of surface singularity.
\end{abstract}
\maketitle

General relativity (GR) predicts that space-time singularities can be formed from regular initial data, e.g. in the gravitational collapse of massive stars and in the early universe.
%
Recently, a modified theory of gravity, the so-called Eddington-inspired Born-Infeld (EiBI) theory was proposed in order to resolve these singularities~\cite{Banados:2010ix}.
The theory was further studied diversely~\cite{Pani:2011mg,Pani:2012qb,DeFelice:2012hq,Pani:2012qd,Cho:2012vg,Cho:2013usa,Avelino:2012ge,Casanellas:2011kf,EscamillaRivera:2012vz,Avelino:2012ue,Liu:2012rc,Delsate:2012ky,Sotiriou:2006qn,Sham:2012qi,Harko:2013wka,Sham:2013sya}.
EiBI gravity is equivalent to GR in vacuum and does not propagate any degrees of freedom other than massless gravitons.
On the other hand, the theory introduces nonlinear couplings to the matter fields~\cite{Pani:2012qb,Delsate:2012ky,Cho:2013usa}, which resolve some of the singularities appearing in GR.
In the presence of a perfect fluid, the big-bang singularity in early universe is replaced by a freezing, exponentially inflating, or a bouncing behavior of the cosmological scale factor, depending on the equation of state (EoS) of the fluid~\cite{Banados:2010ix,Cho:2012vg}.
The gravitational collapse of noninteracting particles does not lead to singular states in the non-relativistic limit~\cite{Pani:2011mg,Pani:2012qb}.
The stability of EiBI compact stars was considered in Ref.~\cite{Sham:2012qi},  where it was shown that the standard results of stellar stability theory still hold.
The structure and physical properties of specific classes of neutron, quark and ``exotic" stars in EiBI gravity were studied~\cite{Harko:2013wka}.
They showed that the EiBI gravity stars are more massive than their GR counterparts.
Constraints on the theory have been considered using solar model~\cite{Casanellas:2011kf} and cosmological observations~\cite{Avelino:2012ge,DeFelice:2012hq}.
A tensor instability of the homogeneous and isotropic universe was found in Ref.~\cite{EscamillaRivera:2012vz}.
In the presence of a scalar field, a regular initial state in EiBI gravity are studied to present the inflation paradigm~\cite{Cho:2013pea,Kim:2013noa}, where the tensor instability was shown to be not a problem.
Metric perturbations on the background of an homogeneous and isotropic universe based on EiBI gravity were studied in Ref.~\cite{Yang:2013hsa,Lagos:2013aua}.

With all these merits, flaws of the theory were also discovered.
In Ref.~\cite{Pani:2012qd}, it was shown that the EiBI theory is reminiscent of Palatini $f(\mathcal{R})$ gravity~\cite{Sotiriou:2008rp} and that it shares the pathology such as the curvature singularity at the surface of polytropic stars.
In Ref.~\cite{Sham:2013sya}, similar singularities were shown to exist when a phase transition happens inside a star.
These observations cast serious doubt on the viability of EiBI gravity.
In this letter, we reexamine the problem by checking the motion of particles at the surface.
In the presence of the curvature singularity, we show that one of the following two things happens depending on the signature of the curvature.
For positive curvatures, the (effective) EoS of particles near the surface is modified so that the singularity is removed.
This happens because the pressure by the particles increases due to rapid oscillations in geodesic deviation equations.
For negative curvatures, on the other hand, the surface becomes unstable.

EiBI gravity is described by the action~\cite{Banados:2010ix}
\begin{eqnarray}\label{maction}
S_{{\rm EiBI}}=\frac{1}{\kappa}\int
d^4x\Big[~\sqrt{-|g_{\mu\nu}+\kappa
\mathcal{R}_{\mu\nu}(\Gamma)|}-\lambda\sqrt{-|g_{\mu\nu}|}~\Big]
+S_M(g,\Phi),
\end{eqnarray}
where $S_M(g,\Phi)$ is the matter action, $\Phi$ generically denotes any matter field, $\mathcal{R}_{\mu\nu}(\Gamma)$ is the Ricci tensor built from the connection $\Gamma$, $|{\cal G}_{\mu\nu}|=\det{\cal G}_{\mu\nu}$,
$\lambda$  is a dimensionless parameter related with the cosmological constant by $\Lambda = (\lambda -1)/\kappa$, $\kappa$ is the extra EiBI parameter which has dimension of length squared, and we set $8\pi G=1=c$.
The equation of motions are obtained by varying the action~\eqref{maction} with respect to the fields $g_{\mu\nu}$ and $\Gamma^\rho_{~\mu\nu}$ respectively,
\begin{eqnarray}
\sqrt{\frac{|q|}{|g|}} q^{\mu\nu} &=& \lambda g^{\mu\nu} -\kappa T^{\mu\nu}, \label{eom1} \\
g_{\mu\nu} &=& g_{\mu\nu} +\kappa R_{\mu\nu}, \label{eom2}
\end{eqnarray}
where $q_{\mu\nu}$ is the auxiliary metric by which the connection $\Gamma^\rho_{~\mu\nu}$  is defined, and $q^{\mu\nu}$ is the matrix inverse of $q_{\mu\nu}$.

Because we are interested in stars in a flat background, we set $\lambda=1$.
We also restrict our interests to the case with positive $\kappa$, because the gravitational collapse does not lead to singularities~\cite{Pani:2011mg}.
We consider a spherically symmetric star made of particles which are effectively described by a perfect fluid with polytropic EoS
\begin{equation} \label{barotropic}
p= p_{\rm poly}\equiv K \rho^\Gamma,
\end{equation}
where $K$ is a constant compensating dimensional difference and $\Gamma$ is a positive dimensionless constant.
The stress-tensor of the fluid is given by
$T^{\mu\nu}=(\rho +p) u^a u^b + p g^{ab}.$
The metric and auxiliary metric for a spherically symmetric star are
\begin{eqnarray}\label{g}
g_{\mu\nu} dx^\mu dx^\nu &=& -e^{\nu(r)} dt^2 +e^{\lambda(r)} dr^2 + R^2 e^{\chi(r)} d\Omega^2 ,\\
q_{\mu\nu}dx^\mu dx^\nu &=& -e^{\beta(r)} dt^2 + e^{\alpha(r)} dr^2+
r^2 d\Omega^2,
\end{eqnarray}
where $\nu(r)$, $\lambda(r)$, $\chi(r)$, $\alpha(r)$, and $\beta(r)$ are arbitrary metric functions of the radial coordinate $r$, $d\Omega^2= d\theta^2+\sin^2\theta d\phi^2$, and $R$ is the radius of the star.
The problem was analyzed precisely in Refs.~\cite{Sham:2012qi,Sham:2013sya,Harko:2013wka} recently.
We follow their definitions for most parts except for the metric function $\chi$, where $f(r)$ was used in place of $R^2e^\chi$ in Ref.~\cite{Harko:2013wka}.
The scalar curvature for the metric $g_{\mu\nu}$ is given by
\begin{eqnarray} \label{Rg}
R_g &=&-e^{-\lambda} \left(\nu''+2\chi''\right) -\frac{e^{-\lambda}}{2} \left[-\nu'\lambda' +\nu'^2 + 2(\nu'-\lambda')\chi'
	+3\chi'^2 \right]+2R^{-2} e^{-\chi},
\end{eqnarray}
where the prime denotes the derivative with respect to $r$.
Outside the star $r>R$, the metric should be the vacuum Schwarzschild solution with
\begin{equation} \label{outer}
e^{\nu} =e^\beta= 1-\frac{2M}{r}= e^{-\lambda}=e^{-\alpha}, \qquad  R^2 e^\chi=r^2,
\end{equation}
where $M$ is the mass of the star.
The continuity equation gives
\begin{eqnarray} \label{continuity}
p ' +\frac{\nu'}{2}(\rho+p) =0.
\end{eqnarray}

Let us describe how the curvature singularity happens at the surface of a star composed of an ideal fluid described by Eq.~\eqref{barotropic} in EiBI gravity.
After solving the equation of motions~\eqref{eom1}, the derivatives of metric functions are given by the sums of the auxiliary metric functions and sources as
\begin{eqnarray} \label{dd}
\lambda' =\alpha'-\frac{a'}{a}-\frac{b'}{b} ,
\qquad
\nu' = \beta' + \frac{a'}{a} -\frac{3b'}{b},
\qquad
\chi' = \frac{2}{r} -\frac{a'}{a} -\frac{b'}{b},
\end{eqnarray}
where
\begin{equation}\label{ab}
a=\sqrt{1+\kappa \rho}, \qquad b=\sqrt{1-\kappa p}.
\end{equation}
After solving Eq.~\eqref{eom2}, the analytic functions $\alpha(r)$ and $\beta(r)$ are determined by two first-order differential equations
\begin{equation} \label{beta1}
(re^{-\alpha})'= 1-\frac{r^2}{2\kappa}\left(2-\frac{3}{ab}+\frac{a}{b^3}\right),
\qquad
(1+ r\beta')e^{-\alpha} = 1-\frac{r^2}{2\kappa}\left(2-\frac{1}{ab}-\frac{a}{b^3}\right).
\end{equation}
From these equations, an EiBI generalization of Tolman-Oppenheimer-Volkoff equation in GR was developed in Refs.~\cite{Sham:2012qi,Harko:2013wka,Sham:2013sya}.
From Eq.~\eqref{eom2}, $\beta''$ can also be expressed as a sum of continuous functions
\begin{equation} \label{beta2}
2\beta'' = \frac{4 e^{\alpha-\beta}}{\kappa}(e^\nu-e^\beta)- \beta'(\beta'-\alpha')-\frac{4\beta'}r.
\end{equation}
Eqs.~\eqref{beta1} and \eqref{beta2} guarantee the continuity of $\beta'$ and $\beta''$ at the star surface, respectively.

Because $\alpha$ and $\beta$ are analytic, possible divergent contributions to the curvature scalar in Eq.~\eqref{Rg} come from the discontinuity of the derivatives of $\rho$ and $p$.
Noting $ \frac{p'}{\rho'} = \frac{dp}{d\rho} = K \Gamma \rho^{\Gamma-1}
 \rightarrow 0$ for $\Gamma>1$,
we find that most singular contributions to Eq.~\eqref{Rg} come from $\rho''$, which are included in $\nu''$ and $\chi''$.
After calculating the explicit form for $\rho^\prime$ and $\rho^{\prime\prime}$ below, one can ensure that $\rho'^2$, $p'^2$, $p^{\prime\prime}$ and $\rho'p'$ do not contain divergent contribution.
The curvature scalar at the stellar surface, as $r\to R$ and $\rho\to 0$, is
\begin{equation}\label{Rg:rho}
R_g \approx -e^{-\lambda} \left(\nu''+2 \chi''\right) = -e^{-\lambda}\left[ \beta''-\frac{4}{r^2} -
	\left(\frac{a'}{a}\right)'-5\left(\frac{b'}{b}\right)'\right]
  \simeq \frac{\kappa e^{-\lambda}}{2} \rho'',
\end{equation}
where we keep only the potentially divergent parts in the first and third equalities and ignore $\kappa\rho$ and $\kappa p$ compared to one.
Assuming the continuity of $\lambda(r)$, we may replace it to its surface value by using $e^{-\lambda(R)} = 1-2M/R$.

With the EoS~\eqref{barotropic}, the continuity equation in Eq.~\eqref{continuity} is integrated to give
\begin{equation}\label{p:barotropic}
\rho  =  \left[ \frac1{K} \exp\left(\frac{\Gamma-1}{2\Gamma}(\nu_0-\nu)\right)-\frac1{K}
    \right]^{1/(\Gamma-1)}.
\end{equation}
For $\Gamma\leq 1$, $p$ vanishes at the places where $\nu \to \infty$.
The energy density and pressure exponentially decrease with $\nu$.
For $\Gamma > 1$, the surface of a star is defined by the place $r=R$ where the pressure vanishes at $\nu(R)=\nu_0$.
Imposing Eqs.~\eqref{barotropic} and \eqref{dd} into Eq.~\eqref{continuity}, we obtain an analytic expression for $\rho'$ as
\begin{equation} \label{rho'}
\rho'(r) = -\beta'(r) g(\rho); \qquad
g(\rho) \equiv \frac{2\rho(K + \rho^{1-\Gamma})} {
    4\Gamma K+\kappa  a^{-2}\rho^{2-\Gamma}
    	+\kappa K H(\rho)},
\end{equation}
where $H(\rho)$ denotes the homogeneous function
\begin{equation}
H(\rho) =\left(a^{-2}+3\Gamma b^{-2}\right)\rho + 3K\Gamma b^{-2} \rho^{\Gamma} .
\end{equation}
Now the second derivative of the energy density becomes
\begin{eqnarray}
\rho''
    &=&-\beta'' g(\rho)
    +\beta'^2 g(\rho)^2 \left( \frac1{\rho}+ \frac{1-\Gamma}
    			{\rho+K\rho^{\Gamma}  }
-\kappa
     \frac{ (2-\Gamma) a^{-2} \rho^{1-\Gamma}-\kappa a^{-4} \rho^{2-\Gamma}+K H'(\rho)  }
   {4\Gamma K+\kappa a^{-2}\rho^{2-\Gamma}+\kappa K H(\rho)} \right), \label{rho''}
\end{eqnarray}
where $H'(\rho)$ denotes the derivative of $H$ with respect to $\rho$,
\begin{equation*}
H'(\rho) = a^{-2}+ 3\Gamma b^{-2}-\kappa a^{-4}\rho
   + 3K \Gamma^2 b^{-2}\rho^{\Gamma-1} \left(1- \frac{\kappa\rho}{b^2}+
    \frac{\kappa K \rho^{\Gamma}}{b^2}\right).
\end{equation*}
Near the surface of the star, we may use $\beta' = 2M/ R^2 \times (1-2M/R)^{-1}$ and $\beta'' = -4M(R-M)/R^2(R-2M)^2$, which are the surface values of the corresponding outer solution.
Taking the $\rho \to 0$ limit, we have
\begin{eqnarray}
R_g \simeq \frac{\kappa}2 \left(1-\frac{2M}R\right)\rho'';
\qquad \rho''\simeq \left\{\begin{array}{ll}
                0, & \qquad 0<\Gamma\leq 1 \\
               \frac{(2-\Gamma)\beta'^2}{(2\Gamma K)^2}
                \left( \rho^{3-2\Gamma}-\frac{\kappa}{4\Gamma K} \rho^{5-3\Gamma}\right),
          & \qquad 1< \Gamma <2 \\
               \frac{-\beta''}{4K+\kappa/2}+
               \frac{ \beta'^2(4K-\kappa)(2K-\kappa)}
               		{2(4K+\kappa/2)^3},
          & \qquad \Gamma =2 \\
               -\frac{2\beta''}{\kappa} +\frac{4\beta'^2}{\kappa}
                \left(1+(2-\Gamma) H'(0)
               -\frac{4\Gamma(\Gamma-2)K}{\kappa^2}
                   \rho^{\Gamma-3}\right), & \qquad \Gamma > 2
             \end{array} \right..  \label{ddrho}
\end{eqnarray}
The curvature scalar diverges for $3/2<\Gamma < 2$ and $2<\Gamma<3$.
This implies that the surface singularity is happening for a star composed of an ideal polytropic fluid.
A crucial part of the results was known in Ref.~\cite{Pani:2012qd}.

An important example of polytropic fluid belonging to this type is the non-relativistic degenerate Fermi gas with $\Gamma=5/3$, which happens for dense Fermi particles when the Fermi energy exceeds by far the temperature.
Examples are the electron gas in metals and in white dwarf stars and the neutron star, whose density is so high that the neutron gas is degenerate.
We derive the EoS for the Fermi gas roughly and then discuss its validity.
At low temperatures and/or high densities we approach the uncertainty principle.
In such a case, the Pauli exclusion principle will cause the pressure to be higher than that inferred by the temperature.
The complete degenerate case happens when two electrons are occupied for each phase space volume.
The average magnitude of the momentum of a Fermi particle is roughly given by
the Fermi momentum,
$
P_{\rm F} = m v_{\rm F}= \hbar\left(3 \pi^2 n\right)^{1/3},
$
where $n$ and $m$ are the number density and mass of a Fermi particle, respectively.
The pressure, total momentum transfer on a unit surface area per unit time, is
\begin{equation} \label{p:Fermi}
p_{\rm F} \approx P_{\rm F} n v_{\rm F} = \frac{\hbar^2}{m} \left(3\pi^2\right)^{2/3} n^{5/3}.
\end{equation}
From this equation, we take the EoS for the degenerate Fermi gas as $ p_{\rm F} = (\rho/K)^{5/3}$ because $n\propto \rho$.

Let us examine how robust the degenerated Fermi gas approximation at the surface of the star.
The approximation holds only when the temperature is smaller than the Fermi energy.
Therefore, the degenerate Fermi gas approximation does not hold for low number densities.
This observation presents one reason to suspect the validity of the polytropic fluid approximation because the surface singularity happens in the low density regime.
However, if a star is very cold, the approximation will be valid even for an extremely low energy density.
Let us check how the geometry reacts on the motions of the particles constituting the fluid.
The EoS~\eqref{p:Fermi} is derived from the motions of Fermi particles in a flat space-time.
On the basis that the energy density and the pressure are local quantities,
the extension to curved space-times is justified by the principle of general covariance.
However, we would like to argue that the pressure may not be local in the presence of a high curvature.
Note that two distant objects, a colliding particle and a probing wall, are necessary
to define the pressure microscopically.
Therefore, if some high curvature effects probe/modify the microscopic structure, the covariant justification for the pressure may not hold,
which is the situation happening near the surface singularity.
To examine the effect, let us study the geodesic deviation equations.
The geodesic deviation $X^a$ along the direction $u^a$  is given by
\begin{eqnarray}
a^0 &=& \frac{2\nu''+\nu'^2-\lambda'\nu'}{4} u^1(X^0 u^1-X^1u^0)
    -\frac{R^2e^{\chi-\lambda}}{4}\nu' \chi '\, u^3(X^0u^3-X^3u^0), \nn \\
a^1 &=& \frac{e^{\nu-\lambda}}{4}( 2\nu''+\nu'^2-\lambda'\nu') u^0
    (X^0 u^1-X^1 u^0)- \frac{R^2e^{\chi-\lambda}}{4}\left[ \lambda'\chi'-
        2\chi''-\chi'^2 \right]\,
         u^3( X^1 u^3- X^3 u^1), \nn \\
a^3 &=& \frac{e^{\nu-\lambda}\nu'\chi'}{4} u^0
    (X^0 u^3-X^3 u^0) +\frac14\left[ \chi'\lambda'
	-2\chi''-\chi'^2\right] u^1(X^1u^3-X^3u^1),
\end{eqnarray}
where we assume $X^2=0=u^2$ without loss of generality because of the spherical symmetry.
Keeping only the dominant parts, we get
\begin{eqnarray}
&&a^0 \simeq\frac{\kappa\rho''}{4} u^1(X^0 u^1-X^1u^0),\qquad
	a^3 \simeq\frac{\kappa\rho''}{4} u^1(X^1u^3-X^3u^1), \nn \\
&&a^1 \simeq \frac{\kappa\rho''e^{\nu-\lambda}}{4}u^0
    (X^0 u^1-X^1 u^0)-\frac{\kappa\rho''\, R^2 e^{\chi-\lambda}}{4}
        \,u^3(u^3 X^1-u^1 X^3),
\end{eqnarray}
where we use $\nu'' \approx \kappa\rho''/2 \approx - \chi ^{\prime\prime}  >0$ for $\Gamma=5/3$.
Consider two geodesics staying away radially so that $X^\mu=(0,X^1,0,0)$.
Consider time evolution only with $u^\mu=(1,0,0,0)$.
Then, the geodesic deviation equation takes the form
\begin{equation}\label{gd:1}
a^1 \approx -\frac{\kappa \rho'' e^{\nu(R)-\lambda(R)}}{4} X^1, \qquad a^0=0=a^3,
\end{equation}
where we set the values of $\nu$ and $\lambda$ to be those at the surface of the star.
Similar expression can be obtained for $a^3$ if we use $u^\mu=(1,1,0,0)$ and $X^\mu=(0,0,0,X^3)$.
These are reminiscent of the Hook's law with frequency
\begin{equation}
f =\frac{1}{2\pi} \sqrt{ \frac{\kappa \rho^{\prime\prime} e^{\nu(R)-\lambda(R)} }{4}}
\approx \frac{\epsilon_0}{4\pi R}
 \left(\frac{M}{R}\right)
 \left(\frac{\sqrt{\kappa}\, \rho^{3/2-\Gamma} }{K}\right) , \label{freq}
\end{equation}
where we use Eqs.~\eqref{outer} and \eqref{ddrho} for the case with $3/2<\Gamma <2$ and each term in the parenthesis is dimensionless.
For the cases with $2<\Gamma <3$, we discuss later.
In the second equality in Eq.~\eqref{freq}, we replace an order one $\Gamma$ dependent term with $\epsilon_0$ to denote that the calculation is approximate.
We should mention that $\epsilon_0$ is an order one dimensionless constant, which depends on the species of particle constituting the fluid.
Following the interpretation of the Hooke's law, any two nearby geodesics (e.g. a wall and a particle) along $X^a$ will cross each other $f$ times irrespective of the distance.

Now, let us calculate the pressure once more as we have done in Eq.~\eqref{p:Fermi}.
Due to the oscillations in the geodesic deviation equation~\eqref{freq}, all Fermi particles inside an oscillation scale collide the wall $f$ times.
The pressure due to the geodesic deviation motion can be estimated as
\begin{equation}
p_{\rm gd} \approx f\ell n P_{\rm F},
\end{equation}
where $\ell$ is the characteristic scale of the oscillations.
Because Eq.~\eqref{freq} is free from scales, $\ell$ appears to be nothing but the star radius, $R$.
Let us try to set $\ell =R$ and examine what happens.
After setting $P_{\rm F} = m v_{\rm F} = A \rho^{1/3} $, where $A$ is a constant, and $n =p_{\rm F}/(v_{\rm F} P_{\rm F})=m K/A^2 \times \rho $, we get
$
p_{\rm gd}\propto \rho^{7/6}.
$
The eventual expression for the pressure will be given by the sum, $p_{\rm F}+p_{\rm gd}$.
In low densities, $p_{\rm gd}$ dominates the pressure over $p_{\rm F}$.
Because $p_{\rm gd}$ acts as if it is that of a polytropic fluid with $\Gamma=7/6<3/2$, the surface singularity disappears.
However, when the curvature becomes too small so that $\rho''$ term is subdominant, there is no reason for the pressure $p_{\rm gd}$ to exist.
Therefore, once again, the EoS will be given by that of the degenerate Fermi gas~\eqref{barotropic} and the surface singularity happens once more.
To avoid this awkward situation, a delicate balance should be taken between the diverging curvature effect and the modification of the EoS so that the curvature is not too small and not too large.
In the presence of such a balance, the curvature must be finite.
In addition, the characteristic scale may not be the radius of the star but be smaller than that.
It is not an easy task to exactly determine how the characteristic size
will be decreased through the balance between the gravity and the Fermi motions.
However, a good measure will be the ratio of the correlation scales between the degenerate Fermion and the gravity per unit time, given by $v_{\rm F}/c$, where we restored $c$ for comparison with $v_{\rm F}$.
One may expect the characteristic length will be decreased by the power of the form, $\ell = R (v_{\rm F}/c)^k$ with $k>0$.
In this letter, we propose $k=1$. 
Explicitly, after setting $\ell =R  v_{\rm F}/c$, we get
\begin{equation} \label{pgd}
p_{\rm gd} = \frac{R f}{c} n v_{\rm F} P_{\rm F} = \frac{R f}{c} p_{\rm F}
    = \epsilon (\rho c^2)^{3/2};
    \qquad \epsilon = \frac{\epsilon_0}{4\pi} \frac{GM}{Rc^2} \sqrt{\frac{8\pi G \kappa }{c^4}}
,
\end{equation}
where we restore $G$ and $c$ with the prescription
$
\kappa \to \frac{8\pi G \kappa}{c^4},~ \frac{M}{R} \rightarrow \frac{GM}{R c^2}, ~ \rho \to \rho c^2,$ and $1/R \rightarrow c/R.
$
Later in this work, we return to use the unit with $8\pi G=1=c$.

Above discussions may also hold for all other matters composing the polytropic fluid with $3/2<\Gamma <2$ because this is due to the geometric effect.
Therefore, for the corresponding particles
we propose a new EoS as
\begin{equation}
p= K \rho^\Gamma+ \epsilon \rho^{3/2} , \qquad \mbox{for}
\quad \frac{3}{2}< \Gamma < 2.   \label{eos2}
\end{equation}
Note that $\epsilon$ decreases with the radius of the star but increases with the mass of the star.
In addition, the correction term vanishes in the GR limit, $\kappa \to 0$.
With the EoS in Eq.~\eqref{eos2}, the second term dominates the pressure when
$
\rho < \rho_c = \left(\epsilon/K\right)^{2/(2\Gamma-3)}.
$
In that regions, the EoS takes after that of the polytropic fluid with $\Gamma= 3/2$, in which case no singularity appears at the star surface.
Higher curvature induces higher pressure.
As the pressure becomes higher, the EoS is modified so that $\Gamma$ decreases.
When $\Gamma$ arrives at $3/2$, the singularity disappears and further modification will not happen.
Therefore, the form in Eq.~\eqref{eos2} appears to be appropriate intuitively once we accept the fact that the curvature modifies the EoS.

Let us calculate the curvature scalar at the surface once more with the new EoS~\eqref{eos2}.
The first derivative $\rho'$ becomes
\begin{equation}\label{rho':new}
\rho' = -\beta' g_m(\rho); \qquad
    g_m(\rho) \equiv \frac{2\rho(K + \rho^{1-\Gamma}+\epsilon
        \rho^{3/2-\Gamma})}
        {4\Gamma K +\kappa a^{-2} \rho^{2-\Gamma}
  +\kappa K H(\rho) +\epsilon \rho^{3/2-\Gamma} G(\rho)},
\end{equation}
where
\begin{eqnarray}
G(\rho) =
6+ \kappa\left[\left(\frac1{a^{2}}+
        \frac9{2b^{2}}\right)\rho
   +\frac{3 K(2\Gamma+3)}{2b^2} \rho^{\Gamma} + \frac{9\epsilon }{2b^2} \rho^{3/2}\right].
\end{eqnarray}
Note that, when $\epsilon=0$, Eq.~\eqref{rho':new} reproduces Eq.~\eqref{rho'}.
The second derivative is given by
\begin{eqnarray}
\rho'' &=& -\beta'' g_m(\rho) +\beta'^2 g_m(\rho)^2
    \left(\frac1\rho \frac{K+(2-\Gamma)\rho^{1-\Gamma}+(5/2-\Gamma)
        \epsilon \rho^{3/2-\Gamma}}{
        K+\rho^{1-\Gamma}+\epsilon \rho^{3/2-\Gamma}
              } \right. \nn \\
    &-&\left.
    \frac{\epsilon(3/2-\Gamma)\rho^{1/2-\Gamma} G(\rho)+
        \epsilon \rho^{3/2-\Gamma} G'(\rho)+
        \kappa\left[(2-\Gamma) a^{-2}\rho^{1-\Gamma} -\kappa a^{-4} \rho^{2-\Gamma} + K H'(\rho)
        \right]}
        {4\Gamma K +\kappa a^{-2} \rho^{2-\Gamma}
  +\kappa K H(\rho) +\epsilon \rho^{3/2-\Gamma} G(\rho)
        }
        \right).
\end{eqnarray}
For $3/2<\Gamma<2$, $\rho'' \simeq \beta'^2/(18\epsilon^2)$.
The curvature scalar at the surface becomes
\begin{equation}\label{Rg:new}
R_g \approx \frac{16\pi^2}{9\epsilon_0^2(1-2M/R)} \frac1{R^2}.
\end{equation}
For an $O(1)$ number $\epsilon_0$, this value is acceptable.
Interestingly, the curvature scalar does not contain $\kappa$ even though it originates from the EiBI gravity.
In GR, the scale of the surface curvature is $R_{GR} \propto \rho \simeq M/R^3$.
$R_g$ is larger than $R_{GR}$ by the factor $R/M$.
For the case of a neutron star, this ratio is not large.

For the matters with $2<\Gamma<3$, the situation is different from the cases with $3/2<\Gamma < 2$.
This is because the signature of the curvature in Eq.~\eqref{ddrho} is negative contrary to the previous case.
Following Eq.~\eqref{gd:1}, two nearby geodesics do not oscillate but repel each other with infinite acceleration.
In the presence of such infinity repulsion, no particle may stay at the surface even in the presence of supporting pressure.
A particle at the surface will be repelled to the center of the star with infinite acceleration.
In the absence of a particle, the radius of the surface decreases, which happens for any value of radius.  
We interpret this as a signature of instability of the surface;
A stable star surface may not exist for the matters.
One may find similar instability happens in the geodesic deviations around a spherically symmetric Newtonian star.
Let us consider the situation that two radially-displaced massive particles fall into the star, which were at rest initially.
Outside the star, the two particles recede relatively because the gravity is stronger inside.
Even in the presence of a constant pressure, their relative distance may not decrease.
On the other hand, inside a static star, they draw nearer because the gravity becomes weaker inside.
The given pressure may keep the particles to stay closer.
Therefore, we may conclude that a star fails to have a stable surface composed of matters with $2<\Gamma<3$, even though such matters can appear inside.

In summary, we studied the phenomena happening at the surface of a star in the EiBI theory, where the star is made of particles which are effectively described by a polytropic fluid with $p\propto \rho^\Gamma$.
For a star composed of a specific kind of ideal-polytropic fluid with $3/2<\Gamma<2$ and $2< \Gamma<3$, a surface singularity is shown to exist.
Noting that the fluid is a macroscopic description of the collective motions of particles, we demonstrated the possibility that a gravitational back-reaction on the matter dynamics modifies the effective descriptions.
The effects are captured by means of the geodesic deviation equation which deals the relative motions between two distant geodesics through the strong gravitational tidal force.
The induced geodesic deviation equation resembles the Hooke's law, where its frequency-squared is proportional to the curvature.
For $3/2<\Gamma<2$, the curvature is positive and the Hooke's law makes a particle collide with a probing wall more often, which modifies the effective EoS.
With the modified EoS, the surface is no longer singular and the size of the curvature at the surface is shown to be acceptable.
On the other hand, for the matters with $2<\Gamma<3$, the curvature is negative and the seeming Hooke's law provides infinite repulsions of the particles.
As a result, a stable star surfaces do not form with the matter.
Note, however, that this result does not imply that the star composed of the corresponding matters does not exist.
Those matters may constitute a core of the star leaving other matters to form the surface.
These two results, lead to the conclusion that there will be absent of surface singularity problem in the EiBI gravity.

Accepting that the EiBI gravity is sound theoretically, one of the most urgent task is to find out observable effects which discriminates the EiBI gravity from GR.
An answer is given in Ref.~\cite{Harko:2013wka}.
There, it was shown that the EiBI stars are more massive than their GR counterparts.
They appeals that some stellar-mass blackhole candidates could be in fact EiBI neutron or quark stars.
However, it is not easy to discriminate this modified-gravity effect from other effects in GR because it is related to the internal structures of the star.
An other prediction is given just above, where it was argued that no star is present with its surface composed of polytropic fluid with $2<\Gamma<3$.
A third observable signature is the relatively large surface curvature (larger than that in GR by the factor $R/M$) for $3/2< \Gamma <2$, which includes neutron stars.
The curvature is independent of $\kappa$ and is almost independent of the mass of the (large) star.
The merit for these two new observations is that the surface curvature is independent of the complex internal structures of the star but is determined by the physics at the surface only.
One may design a scattering experiment of light which grazes the surface of a neutron star.
The light will bears information of the surface curvature, which can be detected from a distant place independently from other information.
Therefore, this fairly large surface curvature will provide an opportunity to distinguish the EiBI theory from GR by means of observation.

In Ref.~\cite{Sham:2012qi}, similar pathology as the surface singularity was shown to happen during the phase transition inside a star.
Because the origin of the singularity is the same as that of the present one, this problem may also be cured with the same prescription.
Similar surface singularities were known to exist in the Palatini $f(\mathcal{R})$ gravity~\cite{Sotiriou:2008rp,Barausse:2008nm}, which origin is known to be the same.
Therefore, we expect that those singularities will be removed with the same way.

\section*{Acknowledgement}
This work was supported by the National Research Foundation of Korea grants funded by the Korea government NRF-2013R1A1A2006548.
The author thanks to Dr. Gungwon Kang for helpful discussions.

\end{document}